\documentclass[twocolumn,showpacs,preprintnumbers,prl,amsmath,amssymb,superscriptaddress ]{revtex4}

\usepackage{amsmath}    
\usepackage{amssymb}
\usepackage{graphicx}   
\usepackage{array}

\newcommand{\BiSe}{Bi$_2$Se$_3${}}
\newcommand{\BiTe}{Bi$_2$Te$_3$~}

\newcommand{\CuBiSe}{Cu$_x$Bi$_2$Se$_3$~}
\newcommand{\Tc}{$T_c$}
\newcommand{\Hc}{$H_{c2}$~}
\newcommand{\Hct}{$H_{c2}(T)$~}

\newcommand{\etal}{{\it et al.}}
\newcommand{\eg}{{\it e.g.}}

\begin{document}

\title{Pressure-Induced Unconventional Superconducting Phase 
\\ in the Topological Insulator Bi$_2$Se$_3$}

\author{Kevin Kirshenbaum}
\author{P. S. Syers}
\author{A. P. Hope}
 \affiliation{Center for Nanophysics and Advanced Materials, Department of Physics, University of Maryland, College Park, MD 20742}
\author{N. P. Butch}
\author{J. R. Jeffries}
\author{S. T. Weir}
 \affiliation{Condensed Matter and Materials Division, Lawrence Livermore National Laboratory, Livermore, CA 94550, USA}	
\author{J. J. Hamlin}
\author{M. B. Maple}
 \affiliation{Department of Physics, University of California, San Diego, La Jolla, California 92093, USA}	
\author{Y. K. Vohra}
 \affiliation{Department of Physics, University of Alabama at Birmingham, Birmingham, AL, 35294}
\author{J. Paglione}
 \email{paglione@umd.edu}
 \affiliation{Center for Nanophysics and Advanced Materials, Department of Physics, University of Maryland, College Park, MD 20742}

\date{\today}

\begin{abstract}

Simultaneous low-temperature electrical resistivity and Hall effect measurements were performed on single-crystalline Bi$_2$Se$_3$ under applied pressures up to 50~GPa. 
As a function of pressure, superconductivity is observed to onset above 11 GPa with a transition temperature \Tc{} and upper critical field \Hc that both increase with pressure up to 30~GPa, where they reach maximum values of 7~K and 4~T, respectively. Upon further pressure increase, \Tc{} remains anomalously constant up to the highest achieved pressure. Conversely, the carrier concentration increases continuously with pressure, including a tenfold increase over the pressure range where \Tc{} remains constant. Together with a quasilinear temperature dependence of \Hc that exceeds the orbital and Pauli limits, the anomalously stagnant pressure dependence of \Tc{} points to an unconventional pressure-induced pairing state in Bi$_2$Se$_3$ that is unique among the superconducting topological insulators.

\end{abstract}


\maketitle


The interplay between superconductivity and topological insulator (TI) surface states has recently received enormous attention due to the observation of the long sought Majorana quasiparticle in InSb nanowires \cite{Mourik} and the promise of realizing topologically protected quantum computation \cite{majorana}.  Characterized by a nontrivial {\it Z}2 band topology with a bulk insulating energy gap that leads to a chiral metallic surface state with spin-momentum locking, TI surface states are analogous to the quantum Hall edge state and arise at the surface of a TI material due to the topological nature of the crossover between a nontrivial bulk insulating gap and the trivial insulating gap of the vacuum \cite{reviews}. 
The use of the proximity effect \cite{Fu096407, Koren224521, Yang104508, Zareapour} to induce superconductivity in \BiSe, the most well studied TI material to date, has had success in coupling these two states but suffers from the presence of bulk conducting states which require gating to realize true TI supercurrents \cite{Cho}.

Theoretically, nontrivial surface Andreev bound states can be directly realized by opening a superconducting energy gap in a bulk conductor \cite{Hsieh107005}, which is why the quest for the topological superconductor is one of the most active areas in condensed-matter physics. Recently, superconductivity has been found in materials with topologically nontrivial band structures, such as in Cu$_x$Bi$_2$Se$_3$ \cite{Hor057001,Wray855,Kriener127004,SasakiZBCP} and YPtBi \cite{Butch220504,heusler}, providing not only intrinsic systems with which to study the interplay between superconductivity and TI states, but also the potential to realize a new class of odd-parity, unconventional superconductivity \cite{Hsieh107005}.

The application of pressure has also uncovered superconductivity in several related materials, such as elemental Bi \cite{Ilina218}, \BiTe \cite{Bi2Te3}, and Bi$_4$Te$_3$ \cite{Jeffries092505}, offering another route to realizing topological superconductivity.  
In this study, we measure transport properties of \BiSe{} over an extended pressure range to investigate the ground state at ultrahigh pressures by using a designer diamond anvil cell capable of measuring both longitudinal and transverse resistivities up to 50~GPa. 
We observe the onset of a superconducting phase above 11 GPa that achieves a maximum transition temperature $T_c = 7$~K above 30~GPa that maintains its value up to the highest pressures achieved in this study. We discuss the implications of an anomalously constant \Tc{} that does not change with pressure, as well as an upper critical field that surpasses both orbital and Pauli limits, in terms of an unconventional superconducting state.


High-quality single crystals of \BiSe{} were grown in excess selenium using the modified Bridgman technique described in detail elsewhere \cite{Butch241301}.  Single-crystal samples---with estimated thickness ($12.5 \pm 2.5$)~$\mu$m and measured carrier concentration $\sim 10^{17}$~cm$^{-3}$---were placed in contact with the electrical microprobes of an eight-probe designer  diamond anvil cell \cite{DAC} configured to allow combinations of both longitudinal and transverse four-wire resistance measurements \cite{SOM}. Pressures were determined from the shift of the ruby fluorescence line \cite{RubyFluorescence}.  Electronic transport measurements were performed at pressures between 4.1 and 50.1 GPa using the standard four-probe technique in both a dilution refrigerator and a pumped $^4$He cryostat, in magnetic fields up to 15~T directed parallel to the $c$-axis of the unpressurized $(R$-$3m)$ crystal structure of \BiSe. Preliminary x-ray diffraction experiments are described.


\begin{figure}[!]
  \includegraphics[width = 3.00in]{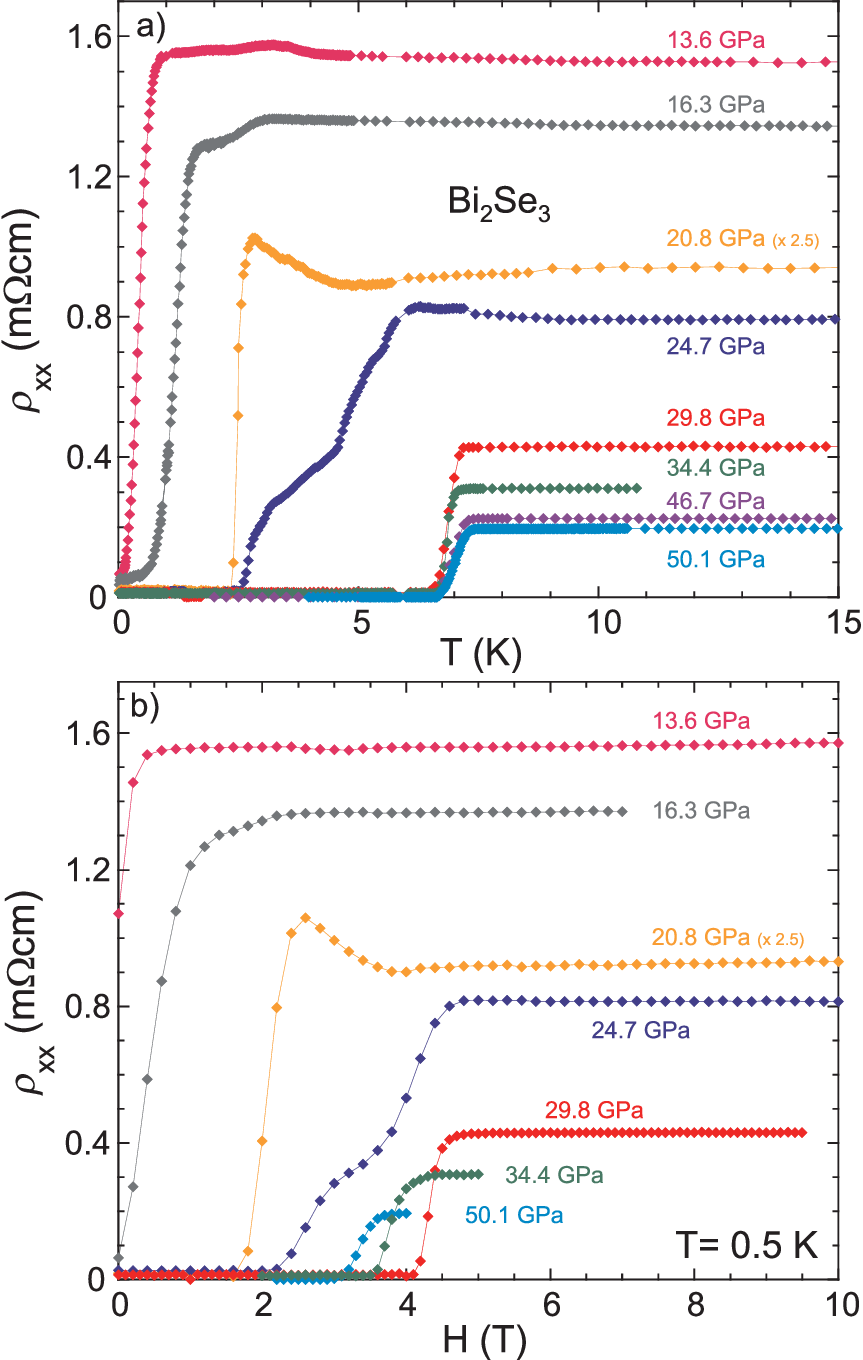}
  \caption{\label{RT} 
  Longitudinal resistivity of \BiSe{} for various applied pressures as a function of a) temperature and b) magnetic field oriented parallel to the crystallographic $c$-axis of the ambient pressure phase, at a fixed temperature of 0.5~K. (Data at 20.8~GPa were obtained with a different lead configuration resulting in larger measurement uncertainty, and are therefore scaled by a factor of 2.5 to match the overall trend reported previously \cite{Hamlin11110098}.) 
  }
\end{figure}

Figure.~\ref{RT} presents a summary of the longitudinal $(\rho_{xx})$ resistivities as a function of both temperature $T$ and magnetic field $H$ measured at pressures above 13 GPa. (Resistivity data measured at lower pressures is presented elsewhere \cite{Hamlin11110098}.)  As shown previously, electrical transport measurements indicate a metallization of \BiSe{} beginning above 8 GPa as revealed by the following: a tenfold decrease in the value of  $\rho$(300 K), a change in the temperature dependence of $\rho(T)$ from semiconducting to metallic conduction, the loss of curvature and development of a linear Hall resistivity $\rho_H(H)$, and the appearance of magnetoresistance that varies with $H^2$ \cite{Hamlin11110098}. Just above this pressure, traces of superconductivity appear in the form of partial resistive transitions, onsetting below 300~mK at 11.9~GPa (not shown) and gradually growing with increasing pressure. Interestingly, the value of carrier density where superconductivity first appears ($\sim$10$^{20}$ cm$^{-3}$) is close to the carrier concentration where superconductivity is seen in \CuBiSe which may indicate that increased carrier concentrations are necessary for superconductivity in \BiSe{} \cite{Kriener127004}.  As shown in Fig.~\ref{RT}(a), a nearly complete resistive transition appears at 13.6 GPa with midpoint transition $T_c = 0.5$~K that gradually increases with increasing pressures up to $\sim$30 GPa. Likewise, as presented in Fig.~\ref{RT}(b), the upper critical field \Hc (defined as the midpoint of the resistive transition in field) also grows with pressure, with a magnetic field dependence very similar in form to the temperature dependence presented in Fig.~\ref{RT}(a), which does not rule out filamentary superconductivity \cite{SOM} but does suggest bulk phase transitions. Also, similar to the pressure evolution of \Tc, \Hc increases monotonically up to 30~GPa, above which both quantities abruptly stop growing and \Tc{} remains strikingly constant at 7~K up to 50.1~GPa. 

A transition temperature that is constant over such a large pressure range is highly anomalous. In conventional phonon-mediated superconductors---like elemental Bi \cite{Ilina218} and the two-band superconductor MgB$_2$ \cite{Tomita092505}---\Tc{} typically decreases with increasing pressure due to phonon stiffening. However when the electronic bandwidth is sensitive to volume change, such as in transition metals, an increase in \Tc{} with pressure is also possible \cite{Hopfield}. These two contrasting pressure-dependent evolutions of \Tc{} are engendered by the {\it{implicit}} dependence of \Tc{} on volume through the phonon cutoff frequency ($\Theta_D$ or $<\omega_{c}>$) and the electronic density of states ($N(E_F)$), as given by the BCS relationship or the McMillan strong-coupling formalism \cite{BCS, McMillan1968}. Thus, for \BiSe, it is possible that these two mechanisms may be balanced so as to produce a pressure-invariant \Tc{} over a wide range of pressure.

\begin{figure}[!]
  \includegraphics[width = 3.00in]{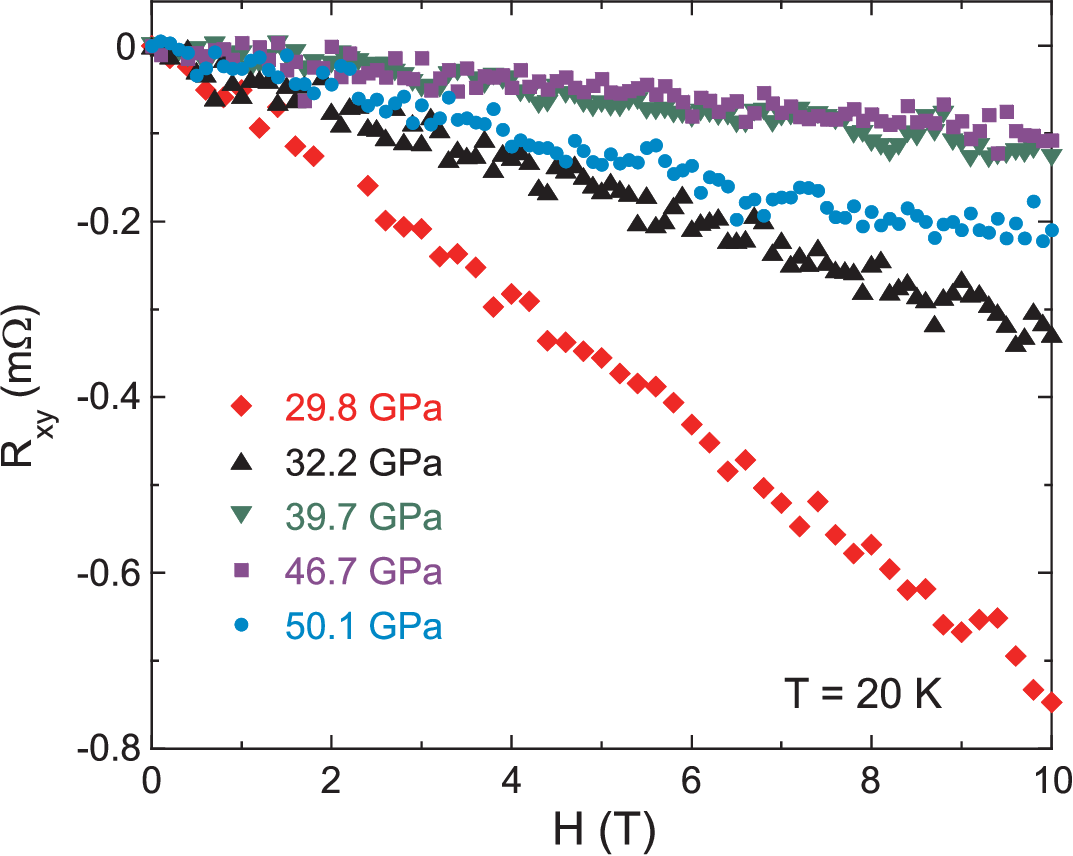}
  \caption{\label{rhoxy} Transverse Hall resistance of \BiSe{} as a function of applied pressure, showing linear behavior with a negative slope indicative of a single, electronlike band. The slope decreases with applied pressure until 46.7 GPa, implying an increasing carrier concentration with pressure; 50.1 GPa presents a larger slope and concordantly smaller carrier concentration (see text for details).}
\end{figure}

\begin{figure}[!]
  \includegraphics[width = 3.25in]{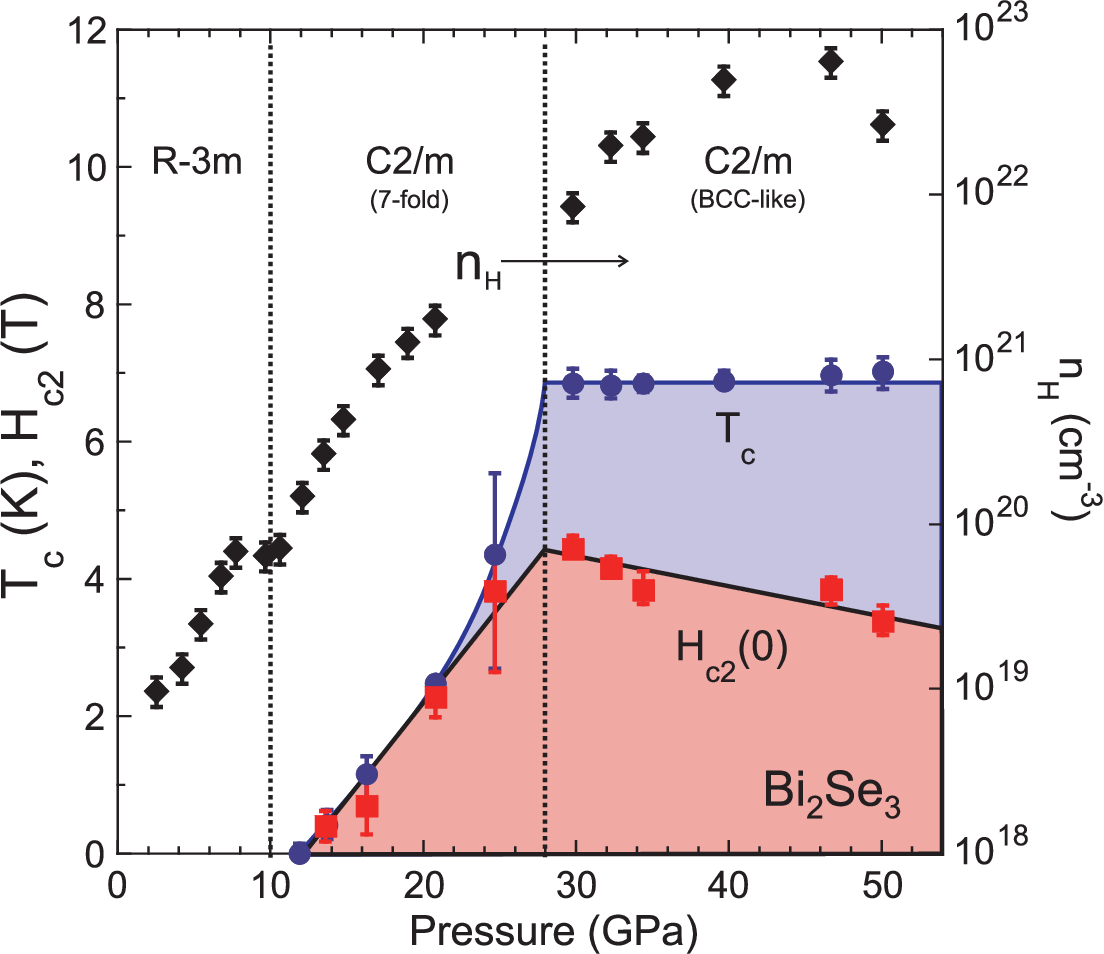}
  \caption{\label{PhaseDiagram} Phase diagram of \BiSe{} showing the evolution of carrier concentration $n_H$ (diamonds), superconducting transition temperature \Tc{} (circles), and upper critical field \Hc at zero temperature (squares) as a function of pressure for fields orientation along the crystallographic $c$-axis of the ambient-pressure structure; $n_H$ data below 21~GPa are from Ref. \onlinecite{Hamlin11110098}. Dotted vertical lines correspond to known structural phase transitions between rhombohedral $(R$-$3m)$ and monoclinic $(C2/m)$ structures near 10 GPa, and a transition to a bcc-like $(C2/m)$ structure near 28 GPa, respectively \cite{SOM,Vilaplana184110,Hamlin11110098}.}
\end{figure}

As shown in Fig.~\ref{rhoxy}, a strong sensitivity of the transverse Hall resistance $R_{xy}$ to pressure suggests that the electronic structure of \BiSe{} indeed undergoes a dramatic change with pressure. A one-band Drude approximation, motivated by the linear field dependence of $R_{xy}$, yields an estimated electron carrier density $n_H$ that increases strongly with increasing pressure, consistent with the increasing metallicity observed in $\rho_{xx}$.  As summarized in Fig.~\ref{PhaseDiagram}, this carrier density increases by over four orders of magnitude over the entire pressure range, suggesting significant changes in the band structure. More surprising, $n_H$ increases by a factor of ten between 30 and 50~GPa, the same range over which \Tc{} remains constant. 
The increasing carrier density with applied pressure suggests an increasing $N(E_F)$, which, by itself, would tend to promote an increasing \Tc. If one assumes a typical pressure-induced phonon stiffening, then the Hall data are at least amenable to a scenario where balanced electronic and phonon contributions lead to the observed pressure invariance of \Tc. However, in either the BCS or the strong-coupling theories, there are other parameters that affect the pressure dependence of \Tc{} \cite{Tomita092505, Hopfield, BCS, McMillan1968}. In the context of phonon-mediated superconductivity, the strikingly pressure-independent value of \Tc{} would necessarily require a fine balance of parameters and an unconventional electronic contribution \cite{SOM}.

Moreover, the arrested evolution of \Tc{} in \BiSe{} is in contrast to that observed in two other closely related compounds where \Tc{} is strongly suppressed with pressure, as found in Bi$_4$Te$_3$ \cite{Jeffries092505} and the closely related TI material \BiTe  \cite{Bi2Te3}. 
Interestingly, \BiSe{} is known to undergo at least two structural transitions under pressure, from the ambient-pressure rhombohedral $(R$-$3m)$ structure to a lower-symmetry monoclinic $(C2/m)$ structure near 10 GPa, and then to an unknown phase above 28 GPa as measured by Raman spectroscopy \cite{Vilaplana184110}.
In both \BiTe and Bi$_4$Te$_3$, superconductivity appears in the monoclinic phase and abruptly strengthens upon crossing a second structural transition into a cubic phase at higher pressures  \cite{Jeffries092505, Vilaplana184110, Vilaplana104112, Einaga092102}. 
Our preliminary x-ray diffraction experiments on \BiSe{} yield similar results, including a structural transition to a sevenfold $(C2/m)$ structure near 10 GPa followed by another transition to a bcc-like $(C2/m)$ structure above 28 GPa \cite{SOM}.
As shown in Fig.~\ref{PhaseDiagram}, the onset of superconductivity in \BiSe{} and its sharp increase to 7~K both coincide with these structural transitions in a manner similar to the other systems, suggesting a close correlation among all of these high-pressure phases.
However, with \Tc{} in \BiTe and Bi$_4$Te$_3$ both exhibiting a notable suppression {\mbox{$dT_c/dP {\sim}$-0.13~K/GPa}} after reaching their maximum values, it is clear that the behavior in \BiSe{} is anomalous.

\begin{figure}[!]
  \includegraphics[width = 3.25in]{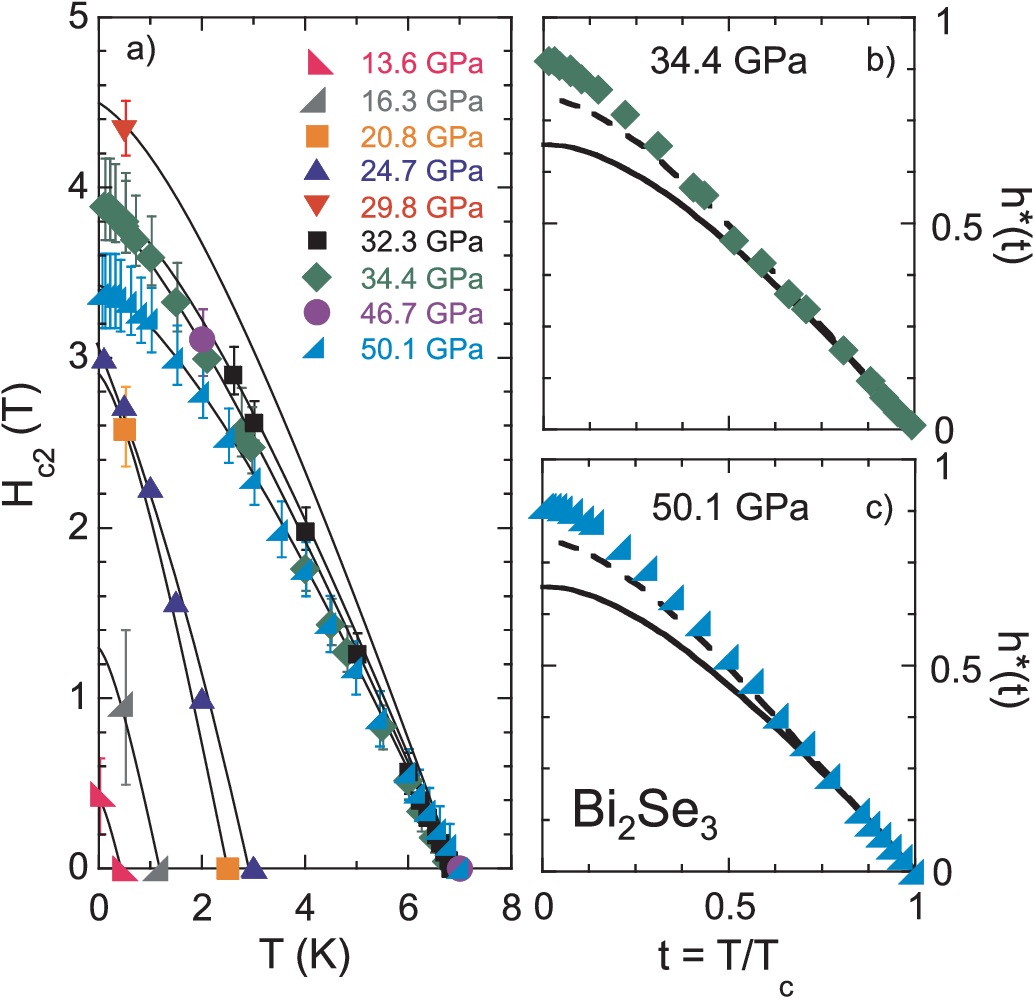}
  \caption{\label{Hc2divTc}  (a) Upper critical field \Hc of \BiSe{} for various pressures up to 50 GPa, with fields applied parallel to the ambient-pressure crystallographic $c$-axis (values determined from 50\% resistive transition, with error bars indicating 10\%-90\% values). Solid lines are guides, but all have the same functional dependence as $H_{c2}(T)$ for 34.4 GPa data. Error bars for 24.7~GPa (not shown for clarity) are  $\pm$1 T.
  Panels (b) and (c) present the reduced upper critical field, $h^*(t)$ with reduced temperature $t=T/T_c$, for applied pressures of 34.4 and 50.1 GPa, respectively. Solid and dashed lines indicate the calculated $h^*(t)$ dependence for orbital limited $s$-wave superconductors \cite{WHHso} and for a polar $p$-wave state \cite{Maki386, Scharnberg5233}, respectively (see text).}
\end{figure}

The unique pressure evolution of \Tc{} in \BiSe{} suggests the presence of a very unconventional superconducting state. This is further evidenced by an anomalous temperature dependence of the upper critical field $H_{c2}(T)$.  
To compare the data to known models, it is useful to calculate the reduced critical field,
$h^*(T) = \frac{H_{c2}(T)}{T_c} / \frac{dH_{c2}(T)}{dT}|_{T=T_c}$,
and compare it to models for orbitally limited $s$-wave \cite{WHHso} and spin-triplet $p$-wave \cite{Maki386, Scharnberg5233} superconductors. As shown in Figs.~\ref{Hc2divTc}b) and c) for 34.4 and 50.1 GPa, respectively, $h^*(T)$ deviates significantly from the expected orbital-limited behavior predicted by the Werthamer-Helfand-Hohenberg (WHH) theory for an $s$-wave superconductor, $H_{c2}^{orb}\simeq 0.7 T_c\times dH_{c2}/dT|_{T=T_c}$ (or $h^*(0)\simeq 0.7$) \cite{WHHso}. This is true through the entire pressure range under study, and is immediately apparent in the observed near-linear temperature dependencies of \Hc shown in Fig.~\ref{Hc2divTc}a). The quasilinear $h^*(T)$ curves in Fig.~\ref{Hc2divTc} are closer in form to that of a $p$-wave superconductor like the heavy-fermion compound UBe$_{13}$ \cite{MapleUBe13}. However, the measured $h^*(0)$ values in \BiSe{} still slightly exceed the maximum value of $h^*(0) \simeq 0.8$ expected for a polar $p$-wave state \cite{Maki386, Scharnberg5233}, further hinting at the unconventional nature of the high-pressure superconducting state of \BiSe.  

To determine the influence of Pauli limiting, we calculate \Hc assuming that both orbital and paramagnetic pair breaking mechanisms are active. The Pauli limiting field $H_{P}$ is determined by the Zeeman energy required to break Cooper pairs and equates to the gap energy $\Delta$ (\eg, $H_P$=1.84$T_c$ for a BCS superconductor) \cite{Clogston266}. 
In the presence of both orbital and Pauli limiting, the expected upper critical field  is modifed to $H_{c2}^\alpha$=$H_{c2}^{orb}$/$\sqrt{1+\alpha^2}$, and determined by the Maki parameter $\alpha \equiv \sqrt{2} H_{c2}^{orb}/H_{P}$ \cite{Maki}. At 34.4~GPa, the calculated values $H_{c2}^{orb}$=3.15~T and $H_{P}$=12.9~T yield $\alpha$=0.346 and an expected modified value $H_{c2}^\alpha$=2.80~T, notably lower than the measured value of 4~T and indicative of an absence of Pauli pair-breaking. A similar case was presented for \Hc measurements of the related superconductors YPtBi \cite{Butch220504, BayYPtBi} and \CuBiSe \cite{Bay057001}, which also both exhibit quasilinear $H_{c2}(T)$ behavior with zero-temperature values exceeding these universal limits. In addition, Bi$_4$Te$_3$ under pressure also exhibits a linear \Hct \cite{Jeffries092505}, presenting an intriguing set of strong spin-orbit-coupled superconducting materials with very similar anomalous features.

While exceeding the WHH limit can be considered a sign of unconventional superconductivity \cite{Clogston266, Scharnberg5233}, other mechanisms should also be considered. 
For instance, Fermi surface topology can enhance the expected WHH limit \cite{Kita224522} as shown in the case of the pyrochlore superconductor KOs$_2$O$_6$ \cite{Shibauchi220506}, although such effects cannot arise from ellipticity alone \cite{KoganProzorovHc2}.  
Strong electron-phonon coupling can also slightly enhance the orbital limit \cite{Bulaevskii11290, Carbotte1027}, although an excessive coupling constant of $\lambda\simeq$4 would be required to explain the observed $h^*(0) \simeq 0.9$. 
Strong spin-orbit scattering was shown early on to greatly reduce the effects of Pauli paramagnetic pair breaking \cite{WHHso}, although a dramatic enhancement is only expected in the limit of infinite scattering strength. 
Finally, multiband superconductivity can also manifest deviations from WHH, as shown for Lu$_2$Fe$_3$Si$_5$ \cite{Nakajima174524}, and calculated for MgB$_2$ \cite{GurevichMultiband} and elemental Bi under pressure \cite{Srinivasan}. While such a case cannot be ruled out for \BiSe{}, the lack of evidence for multiband behavior in the normal-state transport, as evidenced by the linear $R_{xy}$ data in field for the two high-pressure phases, suggests otherwise.

The anomalously large upper critical field that exceeds orbital and Pauli limits and the surprising insensitivity of \Tc{} to pressure point to a unique and unconventional superconducting state in \BiSe. The possibility of this state being topological in nature is an enticing consideration, but requires several as yet unknown criterial to be satisfied. For instance, if band inversion symmetry is present, as well as a Fermi surface that is centered at time-reversal-invariant momenta such that a Dirac-type Hamiltonian describes the band structure, topological superconductivity is indeed probable given a fully gapped pairing symmetry that is odd under spatial inversion \cite{Hsieh107005}. Determination of both crystallographic and electronic structures in the high-pressure phase \cite{SOM} are required to understand the implications for the pairing state and its relation to the ambient pressure topological insulator state. Finally, recent evidence of $s$-wave superconductivity in \CuBiSe \cite{Levy} must be considered in this context.


In conclusion, the metallization of \BiSe{} at high pressures stabilizes a superconducting ground state above 11 GPa that appears to be optimized after a second structural phase transition above 28 GPa. The resulting phase diagram exhibits many similarities to those of other pressure-induced superconducting systems with strong spin-orbit coupling, including the role of structural transitions and the presence of an upper critical field that greatly exceeds the universal predictions for orbital and Pauli pair-breaking. The anomalously large critical fields and the pressure-invariant \Tc{} are incompatible with the expectations of archetypal, phonon-mediated, {\it{s}}-wave superconductors, suggesting the distinct possibility of an unconventional superconducting state in \BiSe.


The authors gratefully acknowledge C.S. Hellberg and I.I. Mazin. 
Work at the University of Maryland was funded by AFOSR-MURI Grant No. FA9550-
09-1-0603.
Portions of this work were performed under LDRD (Tracking Code 11-LW-003).
Lawrence Livermore National Laboratory is operated by Lawrence Livermore
National Security, LLC, for the US Department of Energy (DOE), National
Nuclear Security Administration (NNSA) under Contract No. DE-AC52-07NA27344.
J.J.H. and M.B.M. acknowledge support from the NNSA under the Stewardship Science Academic Alliance program through the U.S. DOE Grant No.DE-52-09NA29459.
Y.K.V. acknowledges support from DOE-NNSA Grant No. DE-FG52-10NA29660.

{\it Note added.}--After submission of this manuscript, we became aware of a similar study by Kong, {\it{et al}} \cite{Kong}.



\clearpage
\bigskip

\begin{widetext}

\begin{center}
{\em \bf \large Supplemental Online Material} \\
\end{center}

\end{widetext}


\section*{The Designer Diamond Anvil Cell}
The designer diamond anvil cell (DAC) for these experiments was composed of an 8-probe designer diamond anvil and a standard diamond anvil, both with culets of approximately 300~$\mu$m in diameter. The microprobes of the designer diamond anvil were tungsten, and they were lithographically deposited to be equally spaced on a 44-$\mu$m diameter circle at the center of the designer anvil culet. The MP35N gasket was pre-indented down to a thickness fo 45 $\mu$m, and the 120-$\mu$m sample chamber was drilled into the pre-indented gasket using an electric discharge machine. Steatite powder was packed into the sample chamber along with a ruby sphere, to be used as a pressure marker. The sample, approximately 10 $\mu$m thick, was placed in contact with the microprobes of the designer anvil, and pressed into the steatite medium upon assembly of the cell. 

Electrical contact is provided by the force of the cell, which physically presses the sample against the electrical contacts (microprobes of the designer anvil). Because of this, each microprobe can make electrical contact with the sample at different pressures, and some microprobes never provide adequate electrical contact. As such, and for these experiments, it was not possible to provide an adequate Hall geometry until $P\ge$29.8 GPa.  Because diamond, owing to the depletion of phonons, becomes a poor thermal conductor at low temperatures, we added a ``thermal strap'' to the DAC in an attempt to mitigate effects associated with poor thermal contact ({\it{e.g.}}, Joule heating). The thermal strap was a thin metal foil that was thermally (but not electrically) connected between the metal gasket and the outside cell body of the DAC.

\begin{figure}[t]
  \includegraphics[scale=0.37]{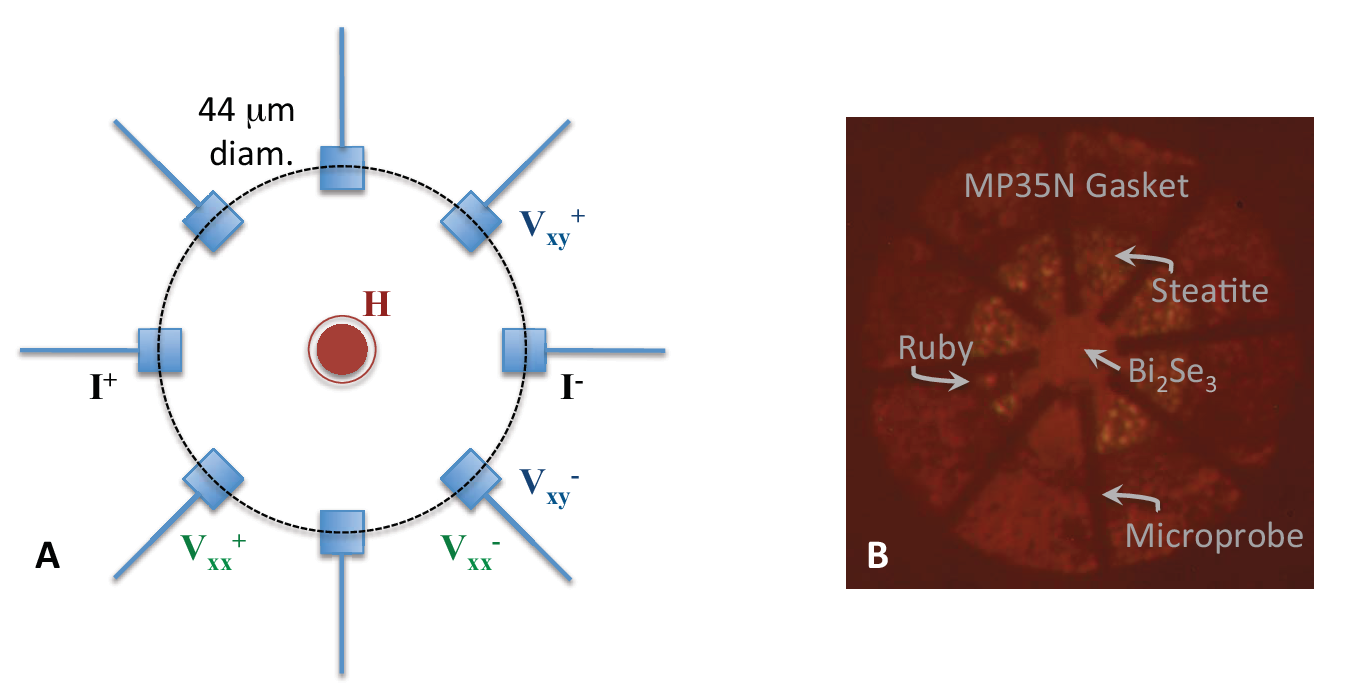}
  \caption{(A) A sketch of the electrical contacts on the culet of the designer diamond. The current and voltage leads used for the longitudinal and transverse electrical transport measurements are labeled. (B) An image of the sample chamber of the assembled designer diamond anvil cell as seen through a red filter looking through the designer diamond.}\label{dDAC} 
\end{figure}

Figure~\ref{dDAC}B is an optical image (taken through a red filter) of the loaded cell looking through the designer diamond. The electrical microprobes, the sample, the ruby pressure marker, the steatite pressure-transmitting medium, and the MP35N gasket material are labeled. 

\section*{Magnetotransport}
The lead configuration for high-pressure measurements is shown in Fig.~\ref{dDAC}A, where the field direction $H$ is out of the plane of the page. For both the longitudinal, $\rho_{xx}$, and transverse, $R_{xy}$, resistance measurements, the current was applied along two opposing leads. $\rho_{xx}$ was measured with the leads labeled $V_{xx}$ and $R_{xy}$ was measured with the leads labeled $V_{xy}$.  $R_{xy}$ was measured for positive and negative fields and the results were symmetrized ({\it{i.e.,}} $[R(+H)-R(-H)]/2$) to obtain the final value of $R_{xy}$ shown in the main text. At low pressures below about 6 GPa---as shown by Hamlin, {\it{et al.}}\cite{Hamlin}---the $R_{xy}$ of \BiSe shows some curvature at higher fields. However, above about 30 GPa (Fig. 2 in main text), $R_{xy}$ appears to be very linear in field. As such, we conservatively use a single-band picture, rather than a compensated multi-band model \cite{Watts}, to extract carrier density from the $R_{xy}$ data. 

\section*{X-ray Diffraction}
Room-temperature, angle-dispersive diffraction patterns were acquired at HPCAT (16 BM-D) of the Advanced Photon Source of Argonne National Laboratory. Conventional DACs were used for these measurements. A neon pressure-transmitting medium was used, and Cu powder was used as the pressure marker. A 10x10 $\mu$m, 32.9 keV (${\lambda}_{inc}$=0.3771 \AA) incident x-ray beam, calibrated with CeO$_2$, was used. 2D diffraction patterns were detected with a Mar345 image plate; exposure times ranged from 60-600~seconds.  2D diffraction patterns were collapsed to 1D intensity versus 2$\Theta$ plots using the program FIT2D\cite{Hammersley1996}.

\begin{figure}[t]
  \includegraphics[scale=0.33]{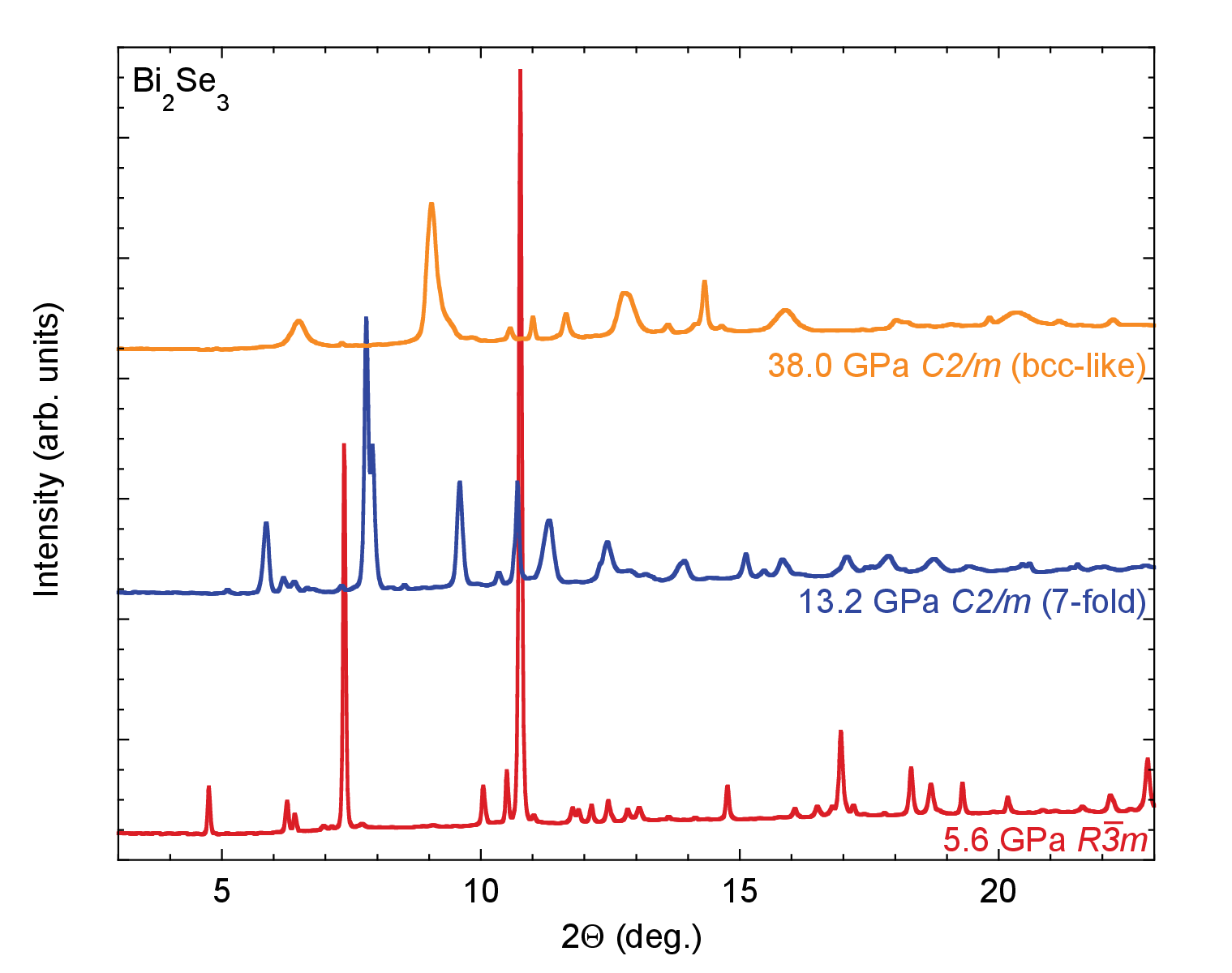}
  \caption{Representative, room-temperature x-ray diffraction patterns for the three different structural variants of \BiSe under pressure. The pressures and space groups are labeled below each diffraction pattern.}\label{XRD} 
\end{figure}

Example x-ray diffraction patterns are shown in Fig.~\ref{XRD}. The patterns show clear, unambiguous changes with applied pressure. The diffraction patterns were indexed and refined using the software program MDI Jade. The results of refinements indicate the following space groups: Bi$_2$Se$_3$-I --- $R{\bar{3}}m$; Bi$_2$Se$_3$-II --- $C2/m$, 7-fold coordinated; and Bi$_2$Se$_3$-III --- $C2/m$, bcc-like coordinated. Phase-II is similar to that reported by Vilaplana, {\it{et al}} \cite{Vilaplana184110}. Our phase-III, however, differs from recent low-temperature results of Kong, {\it{et al.}}, where {\it{C2/c}} and {\it{bcc}} phases are proposed for pressures above 20 and 29 GPa, respectively \cite{Kong}. In our work, the phase transition from Bi$_2$Se$_3$-I to Bi$_2$Se$_3$-II begins near 9.5 GPa and extends just above 10 GPa. The phase transition from Bi$_2$Se$_3$-II to Bi$_2$Se$_3$-III begins near 26.5 GPa and extends just above 30.5 GPa. As the diffraction data was acquired with a highly hydrostatic pressure medium, we expect that the structural transitions may exhibit wider transition ranges in the electrical transport study (above), which used steatite as a solid, pressure-transmitting medium. More details of this structural determination will be included in a forthcoming article \cite{JasonXRD}.

\section*{Pressure Dependence of $T_c$}
The pressure dependence of \Tc can be examined within the scope of a phonon-mediated pairing mechanism. The McMillan strong-coupling formalism provides an excellent starting point to examine the pressure dependence of $T_c$ \cite{McMillan1968, Tomita092505}, which is given by

\begin{equation}
T_c~{\approxeq}~\frac{<\omega>}{1.2} exp \left[ \frac{-1.04(1+{\lambda})}{{\lambda}-{\mu}^*(1+0.62{\lambda})} \right], \label{Eq1}
\end{equation}

\noindent where $<\omega>$ is a characteristic phonon cutoff frequency, $\lambda$ is the electron-phonon coupling strength, and $\mu^*$ is the Coulomb repulsion, which is generally considered to be pressure independent. The pressure-dependent behavior of \Tc can be examined by taking the derivative of Eq.~\ref{Eq1} with respect to volume $V$. This is often done logarithmically, to yield:

\begin{equation}
\frac{dlnT_c}{dlnV}~=~\frac{B}{T_c}\frac{dT_c}{dP}~{\approxeq}~\gamma_G + \Delta \left[ {\frac{dln{\eta}}{dlnV} + 2{\gamma_G}} \right],
\end{equation}

\noindent where $B$ is the bulk modulus, $\gamma_G$ is the Gr{\"{u}}neisen coefficient, $\Delta {\equiv} 1.04{\lambda}[1+0.38{\mu}^*]/[{\lambda}-{\mu}^*(1+0.62{\mu}^*)]$, and $\eta$ is the Hopfield parameter \cite{Hopfield}. The Hopfield parameter itself can be formalized as $\eta=N(E_F)<I^2>$, where $N(E_F)$ is the density of states and $<I^2>$ is an electron-ion matrix element \cite{Medvedeva}. For {\it{s}}- or {\it{p}}-electron systems, the volume-dependent derivative of the Hopfield parameter is generally estimated to be about -1 \cite{Tomita092505}.

Using the bulk modulus ($B$=70 GPa, from x-ray diffraction measurements) of Bi$_2$Se$_3$-III at high pressures, we can examine the values of $\gamma_G$ and $\lambda$ that {\it{could}}, in principle, produce a pressure-invariant $T_c$ as we observe. Setting $\lambda$=1.5 to its highest allowable value for the McMillan formula, we can estimate a $\gamma_G$=1, which is  somewhat low as compared to other materials. Weaker coupling strengths require even smaller values of $\gamma_G$, which are probably physically unrealistic. It should be noted, however, that we do not have any measurements of $\gamma_G$ at high pressures. This would require a measurement of phonons under pressure or measurements of specific heat and thermal expansion at high pressures. 

If one uses the $d$-electron expectations for the pressure-variation of the Hopfield parameter dln$\eta$/dlnV~=~-3.5, then it is possible to find values of $\lambda$ and $\gamma_G$ that fall into ``reasonable'' ranges for these parameters. However, there is, as yet, no justification for expecting a large volume dependence on the Hopfield parameter outside of the realm of usual $p$-electron systems. A quantitative evaluation of the volume dependence of the Hopfield parameter would require electronic structure calculations to accurately correlate the observed changes in the carrier density with changes in the density of states and to compute the electron-ion matrix element. This analysis points to the unconventional nature of the superconducting state of \BiSe under pressure: for phonon-mediated superconductivity to exist, \BiSe must have an unprecedented pressure dependence of its electronic component of superconductivity.

\section*{Filamentary vs. Bulk Superconductivity}
A common concern in electrical transport measurements is the difficulty in determining filamentary versus bulk superconductivity. The high-pressure measurements reported in the main body of this manuscript are performed only with electrical transport, and there is no complimentary ``bulk'' measurement. Nonetheless, the pressure- and field-dependent behavior of $T_c$ in Bi$_2$Se$_3$ suggest that the superconducting state is bulk rather than filamentary.

At the extreme pressure achieved in this experiment, some pressure inhomogeneities are expected in the sample chamber of the DAC. Typically, pressure gradients result in broadening of the superconducting transitions (in our case, at high pressure we have transitions approximately 0.5 K wide) as opposed to a ``shorting out'' of a portion of the sample. However, if the sample were composed of only a few individual, small filaments heterogeneously distributed through the sample, then one might expect to see multiple transitions, where each transition would occur at the local pressure that it experienced. The data (Fig. 1 of main text) are clearly not in favor of distinct, separate superconducting transitions.

At some concentration of filaments, however, it would be difficult to tell the difference (due to the small sample size over which gradients exist) between multiple transitions and a single, broad transition. If the superconducting state was conventional, or if it behaved similar to the Bi-Te analogues, then reproducing the nearly flat pressure dependence of $T_c$ in Bi$_2$Se$_3$ with filamentary superconductivity would require a very special configuration of pressure gradients. From 30-50 GPa, the pressure gradients would have to mimic an identical average pressure without significant broadening of the transition. Furthermore, given that we do not observe multiple transitions in the electrical resistivity, the gradients would have to access various pressures that all lie within about 0.5 K of one another. From this, it would seem that the small pressure dependence of $T_c$ from 30-50 GPa is unlikely to be due to filamentary superconductivity. Furthermore, the relatively large upper critical fields also suggest that the experiments are not probing small filaments of superconductivity. Of course the bulk nature of the superconducting state cannot be irrefutably examined with electrical resistivity alone.

\end{document}